\documentclass[prb,twocolumn,showpacs,amsmath,amssymb]{revtex4}

\usepackage{graphicx}
\usepackage{dcolumn}
\usepackage{bm}
\usepackage{textcomp}

\begin{document}


\title{Comment on "Dielectric behavior of paraelectric KTaO$_{3}$, CaTiO$_{3}$, and (Ln$_{1/2}$Na$_{1/2}$)TiO$_{3}$ under a dc electric field"}

\author{A. Pashkin, V. \v{Z}elezn\'y, M. Savinov and J. Petzelt}
 \affiliation{Institute of Physics, Academy of Sciences of the Czech Republic\\%
Na Slovance 2, 182 21 Prague 8, Czech Republic}

\begin{abstract}
Chen Ang, Bhalla and Cross [Phys. Rev. B \textbf{64}, 184104
(2001)] have studied the low-frequency (20~Hz -- 100~kHz)
dielectric dispersion of the KTaO$_3$ crystal under a dc electric
field. Performing fits of the electric field dependence they came
to conclusion that an appreciable contribution to the dielectric
permittivity originates from polar clusters. In this Comment we
show that the dielectric permittivity at low frequencies (100~Hz
-- 1~MHz) equals to that in the THz region, close below the polar
phonon response. This excludes the possibility of any appreciable
dielectric dispersion due to polar clusters. In addition, we
demonstrate that correct treatment using
Landau-Ginzburg-Devonshire theory allows to fit the electric field
dependence of the dielectric constant without assuming any
polarization mechanism besides the polar phonon modes.
\end{abstract}

\maketitle

The family of incipient ferroelectric materials with perovskite
structure has attracted a great attention due to its high value of
dielectric constant, low dielectric losses and high degree of
tunability by the external electric field and temperature. The
best known members of this family are SrTiO$_{3}$ (STO),
KTaO$_{3}$ (KTO)\cite{Samara01} and CaTiO$_{3}$
(CTO)\cite{Lemanov99} which are also called "quantum
paraelectrics". Microwave measurements of these materials by
Rupprecht and Bell\cite{Rupprecht64} demonstrated that there is no
sign of appreciable dielectric dispersion up to the GHz range.
Therefore it is considered that dielectric properties of pure
incipient ferroelectrics are determined by the contribution of the
lowest polar-phonon mode (the soft-mode) whose frequency is
decreasing on cooling roughly according to the Cochran law.
Recently, combining the far infrared (FIR) spectroscopy with
low-frequency (LF) dielectric measurements some of present authors
confirmed the absence of appreciable dielectric dispersion below
the soft-mode response in STO ceramics\cite{Petzelt01} and CTO
single crystals\cite{Zelezny02}.\par Chen Ang, Bhalla and Cross in
their paper\cite{Chenang01} investigated the dielectric response
of KTO, CTO and (L$n_{1/2}$Na$_{1/2}$)TiO$_{3}$ (L\textit{n} = La
and Nd) (LNTO) under a strong dc electric field (up to 60~kV/cm).
To describe the field dependence of the permittivity, they
attempted to use an equation which is a simple polynomial
expansion by even powers of electric field up to the fourth order.
They were able to describe the field dependence in CTO and LNTO,
but failed to perform a satisfactory fit for KTO. This fact
brought the authors to conclusion that there should be another
Langevin-type term describing the relaxational contribution of
polar clusters to the total polarization. The final equation was
of the form\cite{Chenang01}

\begin{equation}
\label{eq:ang}
\epsilon(E)=\epsilon_{1}-\epsilon_{2}E^{2}+\epsilon_{3}E^{4}+[P_{r}x/\epsilon(0)][\cosh(Ex)]^{-2}
\end{equation}

where $x=P_{r}L^{3}E/2k_{B}T$, $P_{r}$ is the effective
polarization of the polar clusters with the size \textit{L} and
$k_{B}$ is the Boltzmann's constant. Using Eq.~\ref{eq:ang} to fit
the measured field dependence of the permittivity, they obtained
the following parameters of polar clusters:
$P_r=0.4$~$\mu$C/cm$^{2}$, \textit{L} = 4.7 -- 6~nm and
$\Delta\epsilon_{cluster}$ = 656 (at 14~K). This would mean that
23\% of the total dielectric response at low frequencies are due
to some relaxational dispersion of polar clusters below the
soft-mode response.\par In the following we will show that our
recent FIR and LF dielectric measurements of KTO exclude the
possibility of any appreciable contribution to the LF permittivity
besides the contribution of the soft phonon mode. Moreover we will
demonstrate that a correct treatment using the
Landau-Ginzburg-Devonshire thermodynamic theory allows to describe
the field dependence of the dielectric constant in KTO without
using additional Langevin-type terms.\par

Our measurements have been performed on nominally pure KTO single
crystals\cite{Trepakov00}. The samples were prepared in the form
of thin $<001>$-oriented plane-parallel plates with dimensions of
$7\times7\times0.2$ mm$^3$. LF dielectric measurements were
carried out using a HP 4192A impedance analyzer in the frequency
range 100~Hz -- 1~MHz with gold electrodes deposited on the main
surfaces of the sample. The FIR reflectivity was measured with
Bruker IFS 113v Fourier Transform spectrometer in the range 30 --
650~cm$^{-1}$. The complex dielectric function in the frequency
range 5 -- 30 cm$^{-1}$ was obtained by means of time-domain THz
transmission spectroscopy on a 65~\textmu m thick plane-parallel
sample. The frequencies of the three polar phonon modes and their
contribution to the dielectric function were obtained from the fit
to the FIR reflectivity using the factorized form of the model
dielectric function combined with the directly measured values of
the THz complex permittivity\cite{Kamba03} . The temperature
dependence of the soft-mode frequency was found to be in perfect
agreement with the hyper-Raman data\cite{Vogt95}.\par

\begin{figure}
\includegraphics [width=0.9\columnwidth]
    {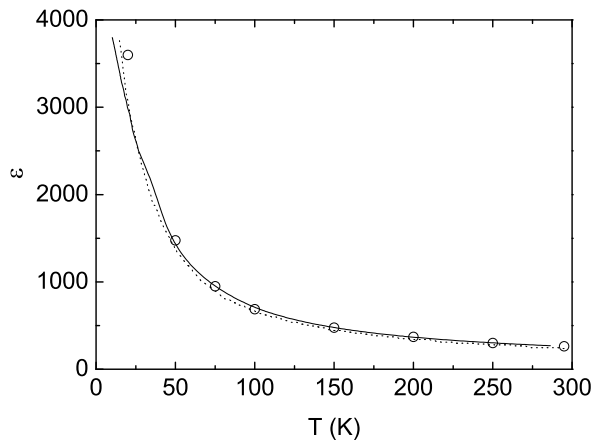}
\caption{Permittivity of KTaO$_3$ crystal at 43.5~kHz (full line),
LF permittivity data taken from Ref.~\onlinecite{Chenang01}
(dotted line) and contribution of the polar-phonon modes (open
circles).\label{fig:compare}}
\end{figure}

In Fig.~\ref{fig:compare} the measured permittivity at 43.5~kHz is
compared with the low-frequency limit of our fitted FIR dielectric
function. The permittivity data from Ref.~\onlinecite{Chenang01}
are also added for comparison. One can observe good agreement
between LF permittivities of KTO crystals and between LF and FIR
permittivities in the whole measured temperature range. Somewhat
higher value of FIR permittivity compared to the LF values at 20~K
lies within the instrumental errors. Anyway,
Fig.~\ref{fig:compare} clearly shows that at all measured
temperatures there is no space for any appreciable dispersion
below the soft-mode response down to kHz range.\par

To demonstrate the fitting of the field dependence of
permittivity, following Fleury and Worlock\cite{Fleury68} let us
expand the crystal free energy per unit volume in a power series
of lattice polarization:

\begin{eqnarray}
\label{eq:freeE} G=G_0+ \frac{1}{2}\chi(T)(P_x^2+P_y^2+P_z^2)+
\frac{1}{4}\xi(P_x^4+P_y^4+P_z^4)\nonumber\\+
\frac{1}{2}\xi^{'}(P_x^2P_y^2+P_y^2P_z^2+P_x^2P_z^2)+
\frac{1}{6}\zeta(P_x^6+P_y^6+P_z^6)\nonumber\\+ \rm\ other\
sixth\textendash order\ terms+\dots
\end{eqnarray}

In our case polarization has only one component directed along the
z-axis ($P_x=P_y=0$). Thus we can obtain relation between the
electric field and polarization:

\begin{equation}
\label{eq:efield}
E_z = (\partial G / \partial P_z)_T =
\chi(T)P_z+\xi P_z^3+\zeta P_z^5+\dots
\end{equation}

which is a special case of the general relation for arbitrary
polarization\cite{Fleury68,Lawless78}. Then one can derive the
expression for the inverse susceptibility:

\begin{equation}
\label{eq:inveps} \left(\frac{1}{\epsilon_0
[\epsilon(T,E)-1]}\right)_{zz}=
\partial E_z /\partial P_z = \chi(T)+ 3\xi P_z^2+ 5\zeta
P_z^4+\dots
\end{equation}

\begin{figure}
\includegraphics [width=0.9\columnwidth]
    {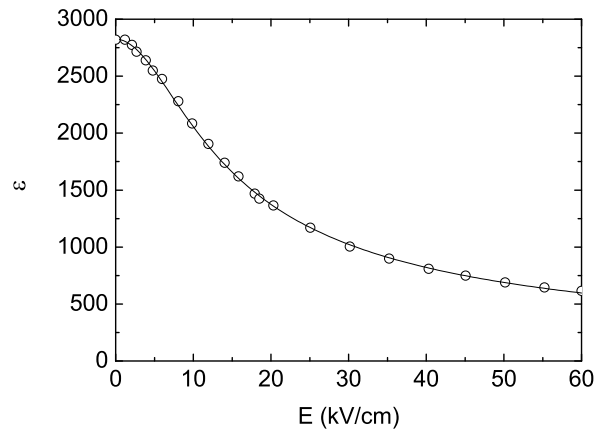}
\caption{Bias electric-field dependence of the KTaO$_3$
permittivity at 14~K taken from Ref.~\onlinecite{Chenang01} (the
first run). Full line corresponds to the fit using
Eq.~\ref{eq:efield} and \ref{eq:inveps}.\label{fig:fit}}
\end{figure}

Here $\chi(T)=1/\epsilon_0[\epsilon(T,0)-1]$ is the inverse
susceptibility in the zero bias field and $\epsilon_0=8.85\cdot
10^{-12}$~F/m is the permittivity of vacuum. To calculate the
permittivity under dc bias field, one has to use
Eq.~\ref{eq:efield} to calculate the polarization and then insert
it into Eq.~\ref{eq:inveps}. The calculations should be performed
numerically because Eq.~\ref{eq:efield} is a \textit{nonlinear}
polynomial equation of the 5-th order. Therefore the analytical
expression given in Ref.~\onlinecite{Chenang01} cannot be valid
for highly anharmonic crystals and can be applied only in the weak
field approximation. This approximation apparently holds in the
case of CTO and LNTO because of their weak nonlinearity for bias
fields $E < 60$~kV/cm. However, in the case of KTO the lattice
anharmonicity is essential. The fit using Eq.~\ref{eq:efield} and
\ref{eq:inveps} of KTO permittivity under the dc electric field at
14~K taken from Ref.~\onlinecite{Chenang01} (the first run of
measurements) is shown in Fig.~\ref{fig:fit}. The fit uses two
independent parameters $\xi=10^{10}$ and $\zeta=10^{11}$ in SI
units and the value $\epsilon$(14 K, 0) = 2821 taken from
Ref.~\onlinecite{Chenang01}. The fitted nonlinear coefficients are
in reasonable agreement with the previously published data (see
Ref.~\onlinecite{Fleury68} and references therein).
Fig.~\ref{fig:fit} demonstrates that the expansion up to the 6-th
order in Eq.~\ref{eq:freeE} is sufficient to ensure a good
agreement between the theory and experiment.\par The critical
limit for the weak field approximation can be found from the
condition that 4-th and 6-th order terms in Eq.~\ref{eq:freeE} are
small compared to the quadratic term

\begin{equation}
\label{eq:critical} \xi P_{cr}^2+\zeta P_{cr}^4=\lambda \chi,
\end{equation}

where $\lambda \ll 1$ is a dimensionless constant. For the typical
values of nonlinear coefficients\cite{Fleury68} we can neglect the
4-th order term in Eq.~\ref{eq:critical} and using
Eq.~\ref{eq:efield} we obtain

\begin{equation}
\label{eq:crfield} E_{cr}=\chi \sqrt{\frac{\lambda \chi}{\xi}}.
\end{equation}

Using our fit parameters for KTO and $\lambda = 0.1$ we calculate
$E_{cr}\simeq 1.3$~kV/cm. One can see that this value is much
smaller than the highest dc electric field in
Ref.~\onlinecite{Chenang01} (60~kV/cm).\par

It is believed that polar nano-clusters are induced by unavoidable
dipolar impurities in nominally pure KTO\cite{Vogt91}. The
presence of such clusters appears in the LF dielectric spectra of
KTO as a universal low-temperature dielectric
dispersion\cite{Trepakov00} with thermally activated behavior.
Below 30~K this dielectric dispersion shifts below the frequency
range of conventional LF dielectric spectroscopy. Dielectric
spectroscopy can only establish that the contribution to
permittivity of polar clusters is very small comparing to the
phonon contribution above 30~K and cannot be observed in LF
dielectric spectra below 30~K. On the other hand, the polar
clusters can be detected by light scattering as it has been done
in Raman\cite{Uwe86} and hyper-Raman\cite{Vogt91} scattering
experiments. These experiments, however, bring no information
about the dynamics of the clusters because their relaxation is
quasi-static comparing to the frequencies resolved in the light
scattering experiments. The local lattice symmetry breaking due to
the polar clusters relaxes the selection rules and induces
activation of new lines. These changes are in agreement with
existence of the polar clusters and do not contradict with our
dielectric spectroscopy results.\par

Taking into account all the mentioned arguments, we can conclude
that the estimated size of polar clusters found in
Ref.~\onlinecite{Chenang01} does not have any reliable physical
background and its possible correspondence with the Raman
results\cite{Uwe86} is accidental.\par

Analysis of the field dependence of permittivity using equation
similar to Eq.~\ref{eq:ang} was previously applied also to $\rm
Sr_{1-x}Ca_xTiO_3$ system\cite{Dec95,Bianchi95} where, due to the
Ca-doping, appreciable dielectric dispersion indeed exists.
However, the applied electric fields were much smaller ($E <
1$~kV/cm) than those in Ref.~\onlinecite{Chenang01}. Estimation of
the critical electric field from Eq.~\ref{eq:crfield} gives for
STO $E_{cr} \simeq 0.2$~kV/cm (nonlinear parameters are taken form
Ref.~\onlinecite{Fleury68}). Thus even 1~kV/cm field can introduce
some error if one uses Eq.~\ref{eq:ang}. In our opinion, the
correct thermodynamic approach described here is always preferable
for the analysis of nonlinear behavior under strong electric
fields.\par

\vspace{3 mm} This work was supported by the Grant Agency of the
Czech Republic (projects No. 202/01/0612 and 202/02/0238) and
Czech Academy of Sciences (project AVOZ-010-914).

\end{document}